\documentclass[
    ,final            
  ]
  {aipproc}

\layoutstyle{6x9}

\begin{document}

\title{Direct photon interferometry}


\author{D. Peressounko}{
  address={RRC "Kurchatov Institute"}
}

\begin{abstract}
We consider recent developments in the theory of the two-photon
interferometry in ultrarelativistic heavy ion collisions with
emphasis on the difference between photon and hadron
interferometry. We review the available experimental results and
discuss possibilities of measurement of the photon Bose-Einstein
correlations in ongoing and future experiments.
\end{abstract}

\maketitle


Direct photon interferometry is the one of the most interesting
and informative tools for exploring properties of the hot matter,
created in heavy ion collisions. Photons have extremely large free
path length in the hot matter and deliver direct information about
space-time dimensions of the inner hottest part of the collision.
Moreover, the direct photons, emitted at different stages of the
collision, dominate in the direct photon spectrum in different
ranges of transverse momentum, therefore, measuring correlation
radii at different $K_T$ one can extract space-time dimensions of
the system at the different stages of the collision and thus
access the equation of state of the hot matter.

Direct photons contribute only a small fraction of the total
photon yield while the dominant part of inclusive photons comes
from decays of final hadrons, mainly $\pi^0$ and $\eta$ mesons.
Fortunately, the lifetime of these hadrons is extremely large, and
the width of Bose-Einstein correlations between decay photons is
of the order of a few eV, so that it can not be observed and it
does not obscure the direct photon correlations. So, when one
talks about the photon interferometry, one means the correlations
of direct photons.

Technically the interferometry of direct photons in most respects
is similar to the hadron interferometry, but still it has several
specific features, which make it special. These features, which
will be discussed in details below, are related to the following
properties of a photon:
\begin{itemize}
\item
  The penetrating nature of the direct photons. Since direct photons are
  emitted from the central zone of the collision and photons with
  different $K_T$ are emitted at the different stages of the
  collision, photon interferometric correlation radii do not
  follow the $M_T$ scaling, usual for hadrons, but have more complicated shape.
\item
  Zero mass of a photon results in specific interpretation of
  the invariant correlation radius and correlations strength
  parameter, and requires some special one-dimensional parameterization
  of the photon correlations.
\item
  Small proportion of direct photons in the total photon yield
  makes photon correlation strength parameter very small what in turn
  leads to the importance of background photon correlations.
\end{itemize}

\begin{figure}[!t]
\includegraphics[width=0.32\textheight,height=0.32\textheight]{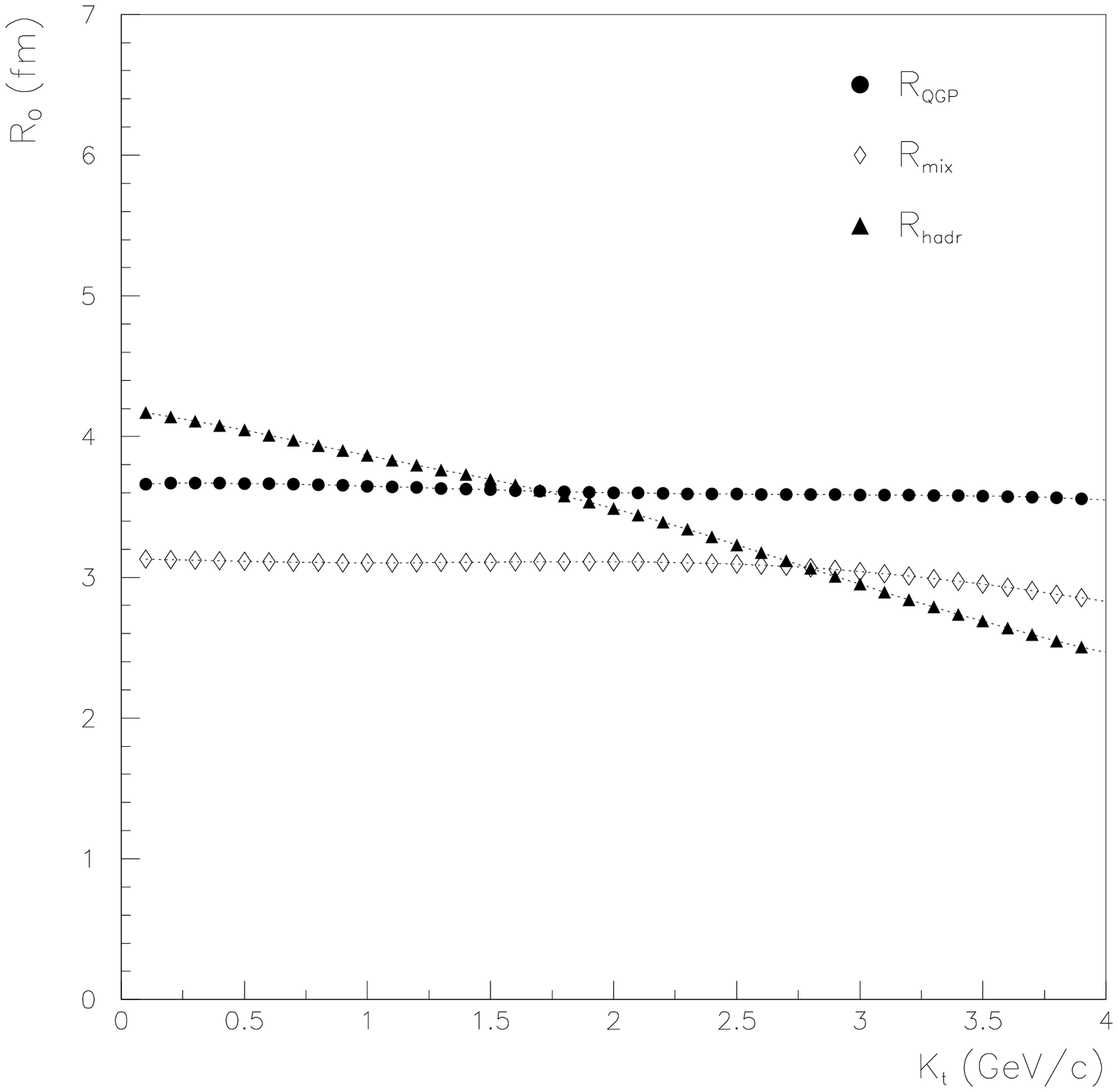}
\includegraphics[width=0.32\textheight,height=0.32\textheight]{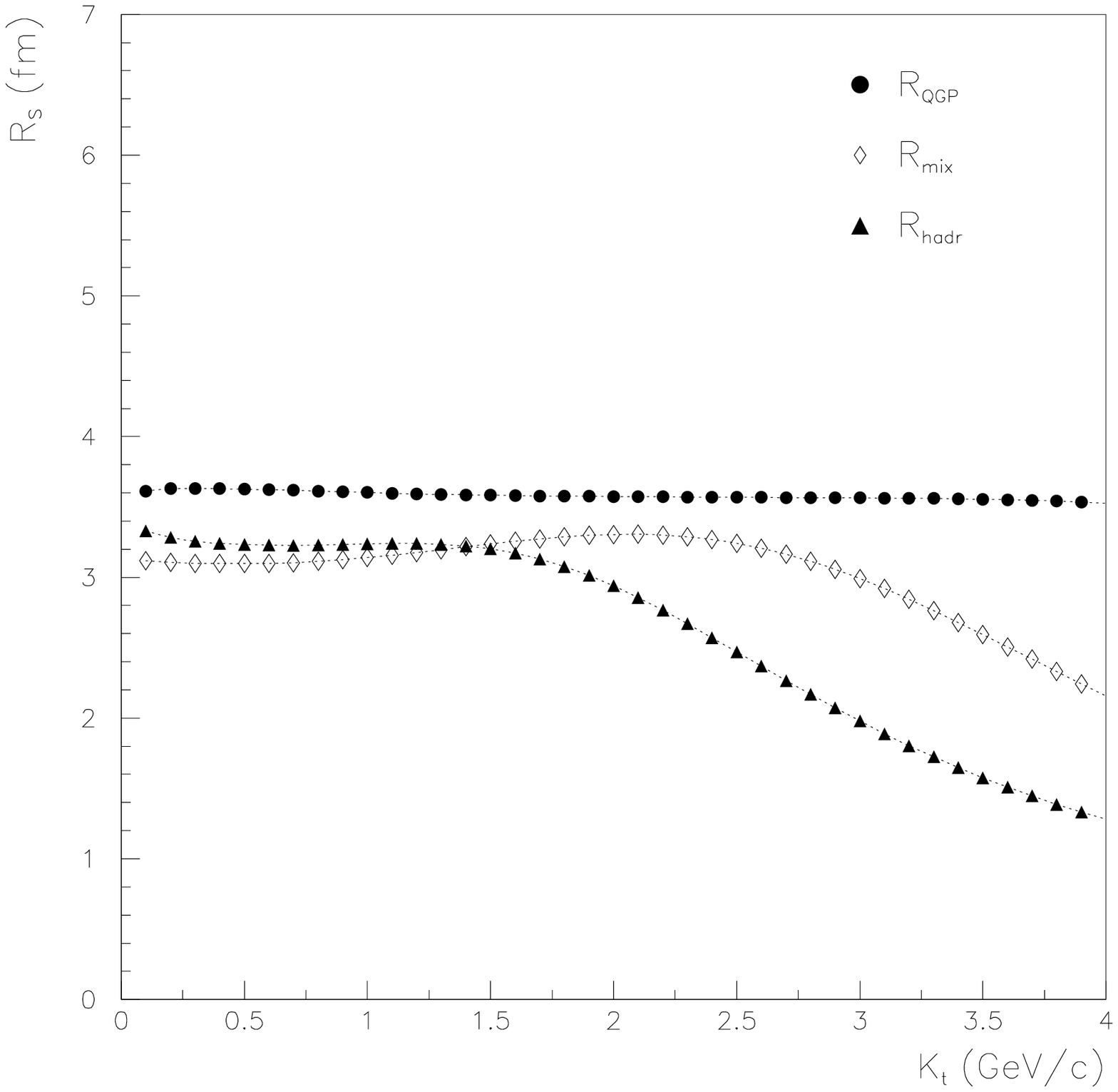}
  \caption{Two-photon correlation radii in Au+Au collisions at
  RHIC energy \cite{Peressounko} for "side" (left plot) and
  "out" (right plot) directions. Contributions of QGP, mixed and
  hadronic phase are shown separately.}
  \label{fig:rad-dec}
\end{figure}

\begin{figure}[!t]
\raisebox{0pt}[0.32\textheight][0pt]{
\includegraphics[width=0.32\textheight,height=0.32\textheight]{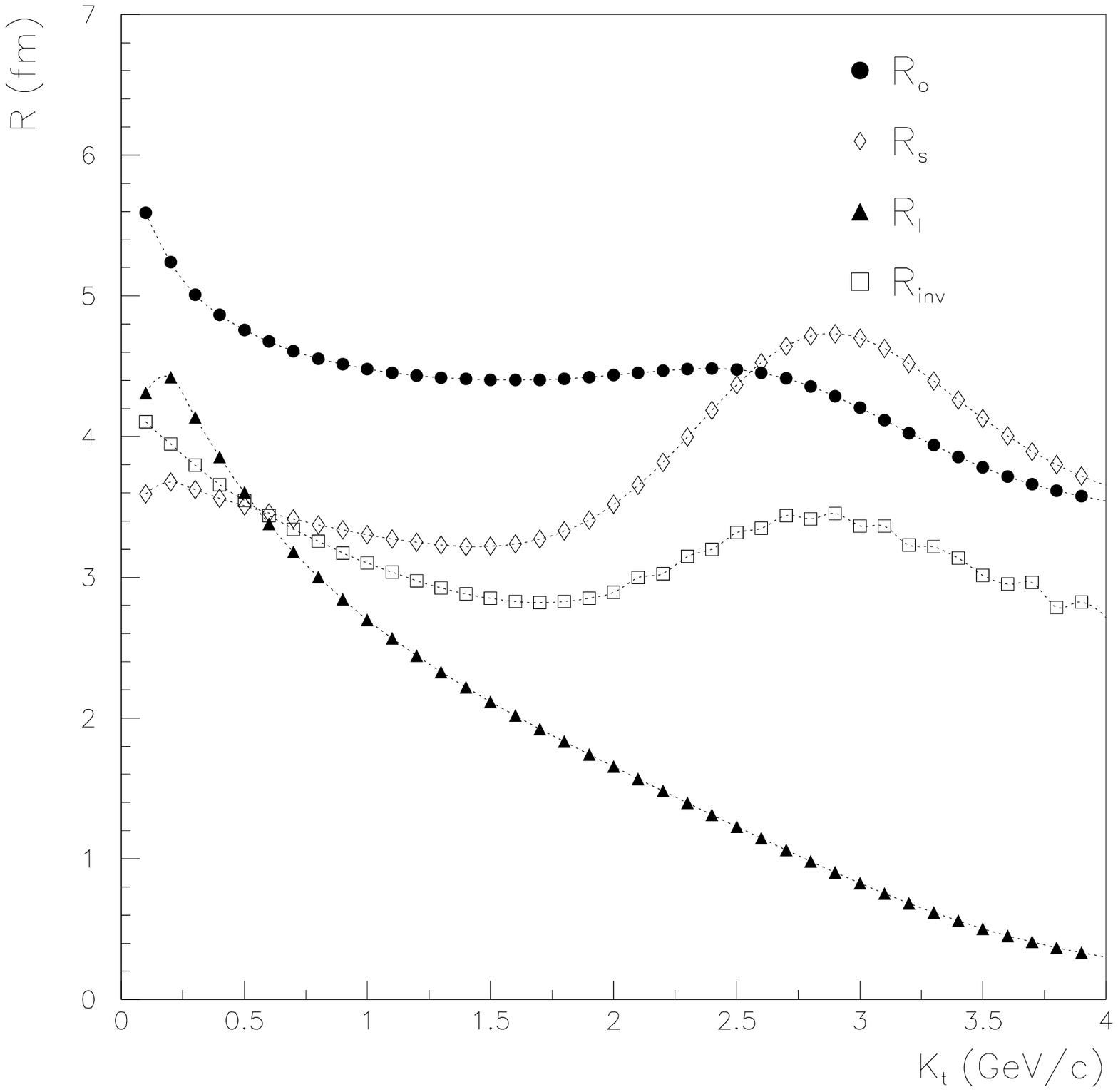}
\includegraphics[width=0.36\textheight,height=0.32\textheight,angle=-90]{fig2b.eps}
}
  \caption{$K_T$ dependence of different components of photon correlation radii in
  Au+Au collisions at RHIC energy, obtained with 2+1 Bjorken hydrodynamics
  \cite{Peressounko} (left plot) and $R_{side}$ dependence within model with
  parameterization with constant acceleration \cite{Renk} (right plot).}
  \label{fig:radii}
\end{figure}

First predictions of the direct photon intensity correlations in
heavy ion collisions have been published long ago
\cite{Makhlin,Srivastava1,Razumov,Peressounko}, while the next
bunch of calculations \cite{Alam-gg,Srivastava2,Renk} appeared
after publishing of the first experimental results on direct
photon interferometry by WA98 collaboration \cite{wa98}. Despite
the large number of publications, up to now there is no agreement
between predictions not only in the absolute values of the
correlation radii of direct photons, but even on the shape of
$K_T$ dependence of the correlation radii. For example, in the
Fig.~\ref{fig:rad-dec} we present predictions of the "out" and
"side" correlation radii of photons, emitted in Au+Au collisions
at RHIC energy \cite{Peressounko}. These predictions are made
within 2+1 Bjorken hydrodynamics with the first order phase
transition. Contributions from the different phases: QGP, mixed
and hadronic phase are shown separately. We find no $K_T$
dependence for photons, emitted from the QGP phase (including pQCD
photons), radii from the mixed phase is constant up to $K_T\sim 3$
GeV where Doppler-shifted contributions from a few accelerated
pieces became important, and photons emitted from the hadronic
phase exhibit some $K_T$ dependence in agreement with the large
collective flow developed in this phase. We compare two
predictions of direct photon correlation radii for central Au+Au
collisions at RHIC energy ($\sqrt{s_{\small NN}}=200$ GeV), in the
Fig.\ref{fig:radii}. In the left plot we present result of the
calculations, made within 2+1 Bjorken hydrodynamics
\cite{Peressounko}. In the right plot we show results obtained
using parameterization of the evolution with a constant
acceleration \cite{Renk}. We find the bump and the region with
$R_s>R_o$ in the first case and the $M_T$ scaling in the second
case. Whether this discrepancy can be attributed to the difference
in evolution or it is related to the details of extraction of
correlation radii is not clear yet.

Strength of direct photon correlations is usually extremely small,
being on the level of tenth of percent, therefore it is difficult
to gather sufficient statistics to construct the full
three-dimensional correlation function. So one has to deal with
averaged one-dimensional distributions. However, for massless
particles the averaging from the full three-dimensional
correlation function to those, depending on $Q_{inv} =
\sqrt{-(k_1-k_2)^2}$ has some unexpected features. To illustrate
this, let us consider a simple toy model: assume, that photons are
emitted from the symmetric Gaussian source with the radii
$R_x=R_y=R_z=R$ and the emission duration $\tau$ so that the full
correlation function of photons is
$$
    C_2(q,K) = 1+\lambda \exp\left ( -q_x^2 R^2 -q_y^2 R^2 -q_z^2 R^2 -q_e^2 \tau^2  \right
    ),\nonumber
$$
where $q$ and $K$ are relative and average momenta of the pair and
$\lambda$ is the correlations strength parameter. To go from the
full three-dimensional to the one-dimensional parameterization, we
have to integrate over the components of the relative momentum
under the condition $\delta(Q_{inv}^2+q^2)$. The result can be
expressed as follows:
\begin{eqnarray}\label{C2-cm}
    C_2(Q_{inv},K)\!\!\!&\!\!\!\!=\!\!\!\!&\!\!\!\frac{1}{4\pi}\int \! C_2(\hat q,\hat K)\, d\Omega \nonumber
    \!=\!\frac{1}{4\pi}\int\!\!\!\left[
    1+\lambda \exp\left(-Q_{inv}^2 R^2
    -4K_T^2\cos^2\!\theta(R^2+\tau^2)\right) \right ]\,
    d\Omega \nonumber \\
&\!\!\!\!=\!\!\!\!& 1 + \lambda_{inv}\exp\left (-Q_{inv}^2
R^2\right ), \nonumber
\end{eqnarray}
where $\hat q$ and $\hat K$ are the relative and average pair
momentum in pair CM frame, the integration $d\Omega$ is done over
directions of the relative momentum and $K_T$ is average
transverse momentum of the pair. We find that the correlation
strength of the one-dimensional correlation function is
considerably reduced:
$$
\lambda_{inv}=\frac{\lambda}{2}\int\!d\cos\!\theta
\exp\left(-4K_T^2\cos^2\!\theta(R^2+\tau^2)\right)=\frac{\sqrt{\pi}\,\,
\rm{erf}(2K_T\sqrt{R^2+\tau^2})}{2K_T\sqrt{R^2+\tau^2}}.
$$
Calculations with the more realistic source demonstrate that the
invariant correlation radius of massless particles is (using
out-side-long three-dimensional radii) an average of the $R_s$ and
$R_l$ correlation radii and almost independent on the $R_o$
component, while the invariant correlation strength parameter
depends on the product $(K_T R_o)$ and is con\-si\-de\-rably
smaller than the three-dimensional correlation strength parameter.
This takes place because $Q_{inv}$ for massless particles depends
mainly on the opening angle and is almost independent on the
$q_{o}$ so we average over this component in the range $|
q_{o}|\le 2\,K_T$.

\begin{figure}
  \includegraphics[width=0.32\textheight,height=0.32\textheight]{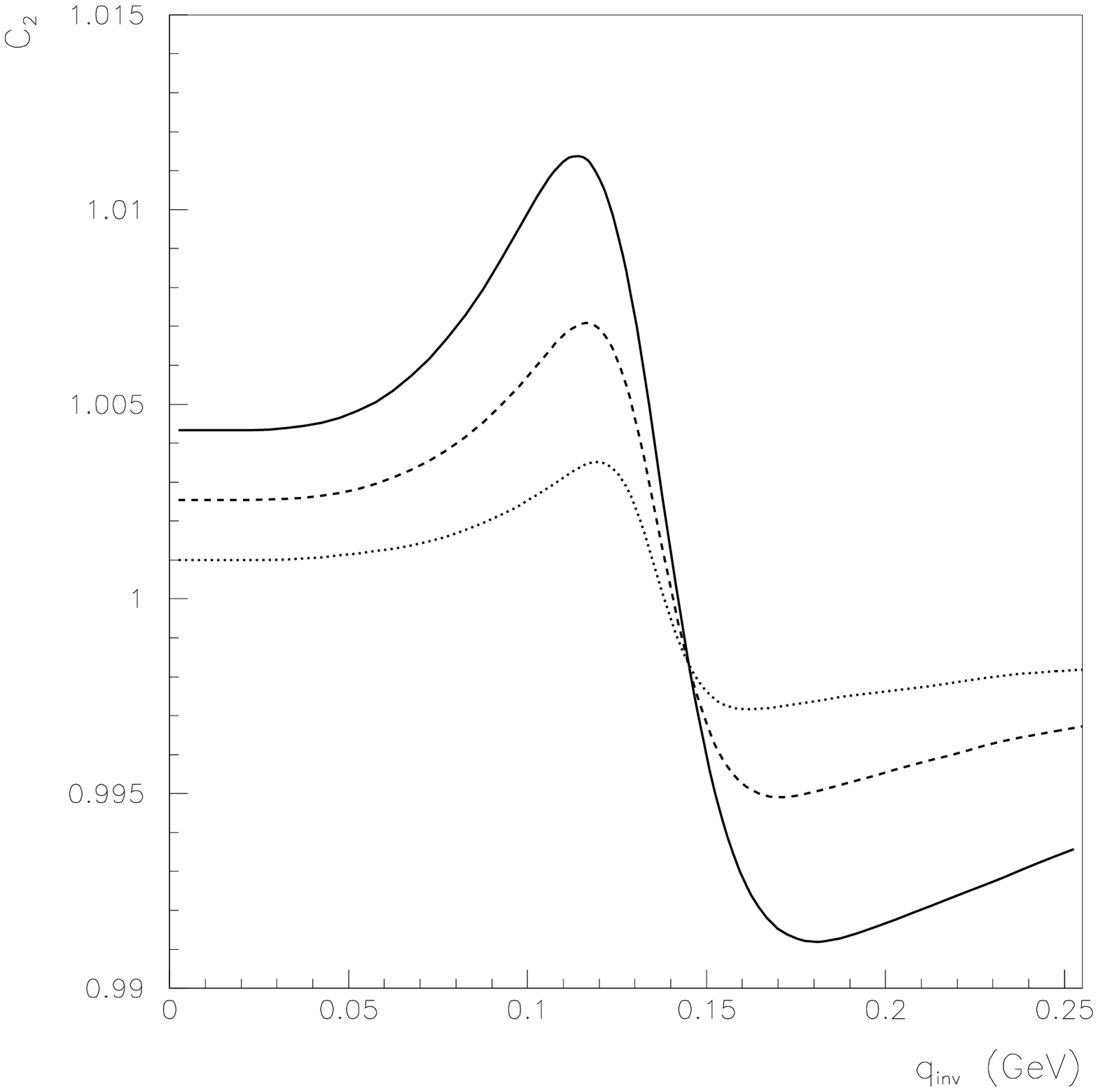} \hfill
  \includegraphics[width=0.32\textheight,height=0.32\textheight]{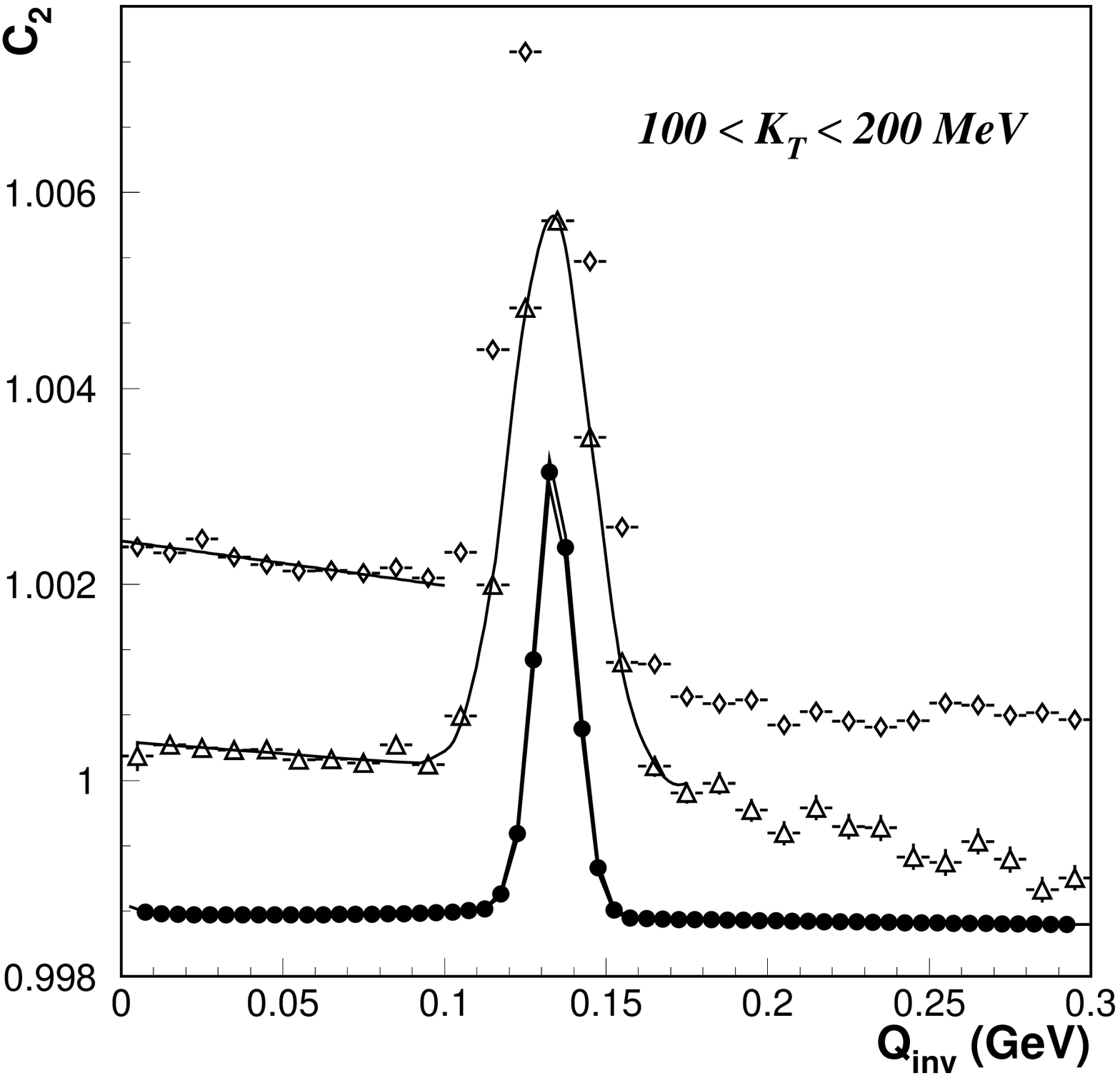}
  \caption{Decay photon background correlations. Left plot is residual
  correlations between products of Bose-Einstein correlated $\pi^0$, calculated
  for 3 different radii of pion correlations: 4, 5 and 6 fm (solid, dashed and
  dotted lines correspondingly)  \cite{Peressounko}.
  Right plot presents result of Monte-Carlo
  simulations of different residual correlations within WA98 acceptance and
  experimental cuts \cite{wa98} due to $\pi^0$ BE correlations (diamonds),
  elliptic flow (triangles) and kinematic correlations (boxes).}
  \label{fig:bg}
\end{figure}

The extremely small strength of the direct photon correlations
leads to the importance of the photon background correlations:
even small correlations between decay photons may completely hide
the direct photon correlations. Since the decay photons originate
in decays of final hadrons, they may carry some residual
correlations. Keeping in mind that the main part of the decay
photons comes from $\pi^0$ decays, one can classify the background
correlations as following: (1) residual correlations between the
decay photons originated from Bose-Einstein correlated neutral
pions; (2) residual correlations between products of kinematically
correlated pions or photons, e.g. photon correlations in the
processes $K^0_S\to 2\pi^0\to 4\gamma$ or $\omega\to\gamma\pi^0\to
3\gamma$; (3) residual correlations due to collective (elliptic)
flow of parent pions. The first point (1) is the most dangerous
since if the shape of these residual correlations will repeat the
shape of Bose-Einstein correlations of the parent pions, this
background will completely hide the direct photon correlations.
Fortunately, this is not the case. One can analytically
demonstrate, that the residual correlations due to pion
Bose-Eeinstein correlations have a characteristic wave-like shape
with the plato at small $Q_{inv}$ (see Fig.~\ref{fig:bg}) and can
be disentangled from the direct photon correlations. Monte-Carlo
simulations made by WA98 collaboration \cite{wa98}
(fig.~\ref{fig:bg}, right plot) and by Utyuzh et al. \cite{Utyuzh}
support this conclusion. The shape of the background correlations
due to the kinematic correlations (case 2) and the elliptic flow
(case 3) strongly depends on the apparatus acceptance and the
experimental cuts used in analysis, but usually it appears as a
long range correlations, as presented in the Fig.~\ref{fig:bg},
right plot. So, the background photon correlations in
nucleus-nucleus collisions can be disentangled from the direct
photon correlations but should be accounted in calculation of the
correlation parameters.

\begin{figure}
\raisebox{0pt}[0.315\textheight][0pt]{
  \includegraphics[width=0.32\textheight,height=0.29\textheight]{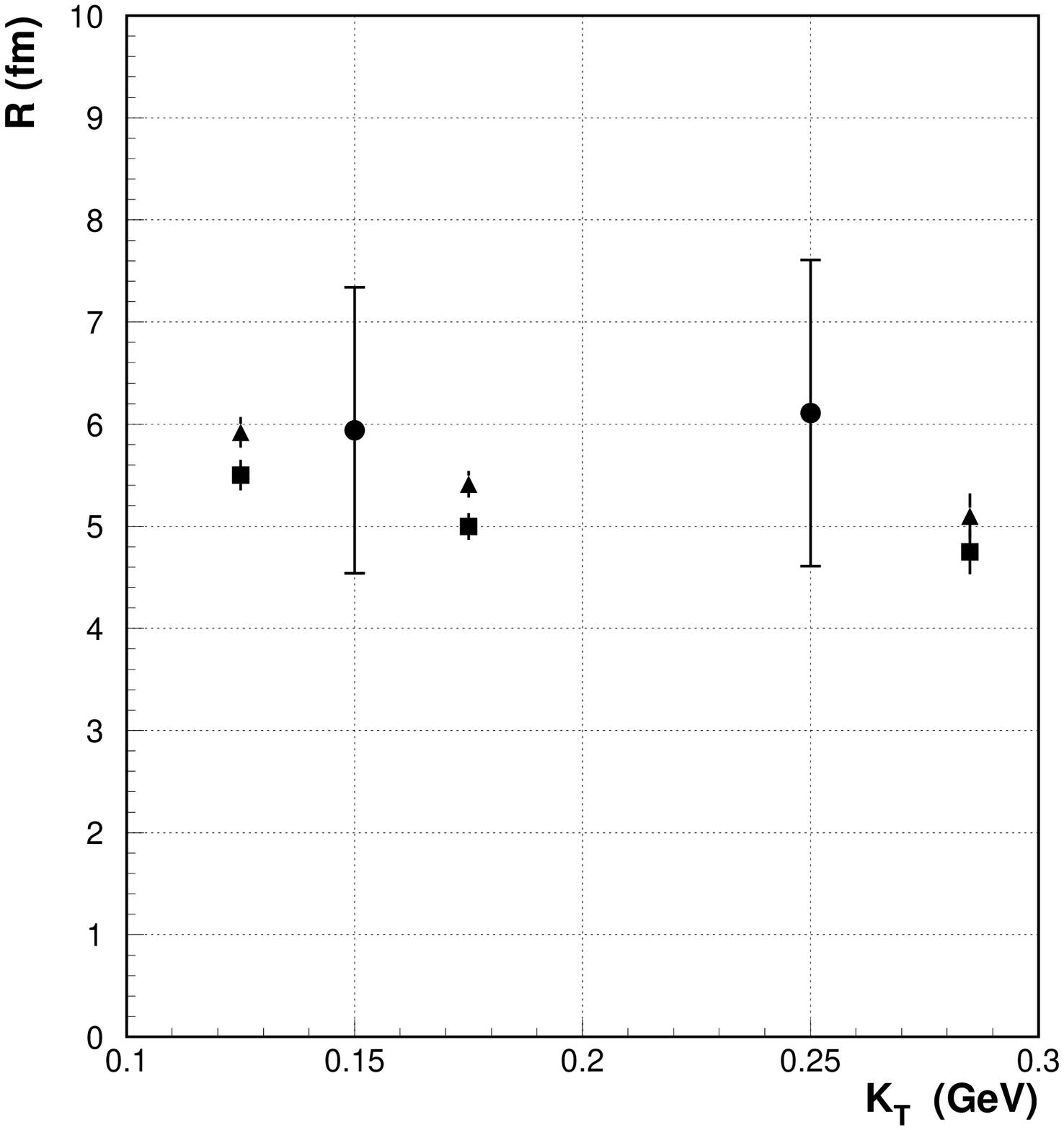} \hfill
  \includegraphics[width=0.32\textheight,height=0.29\textheight]{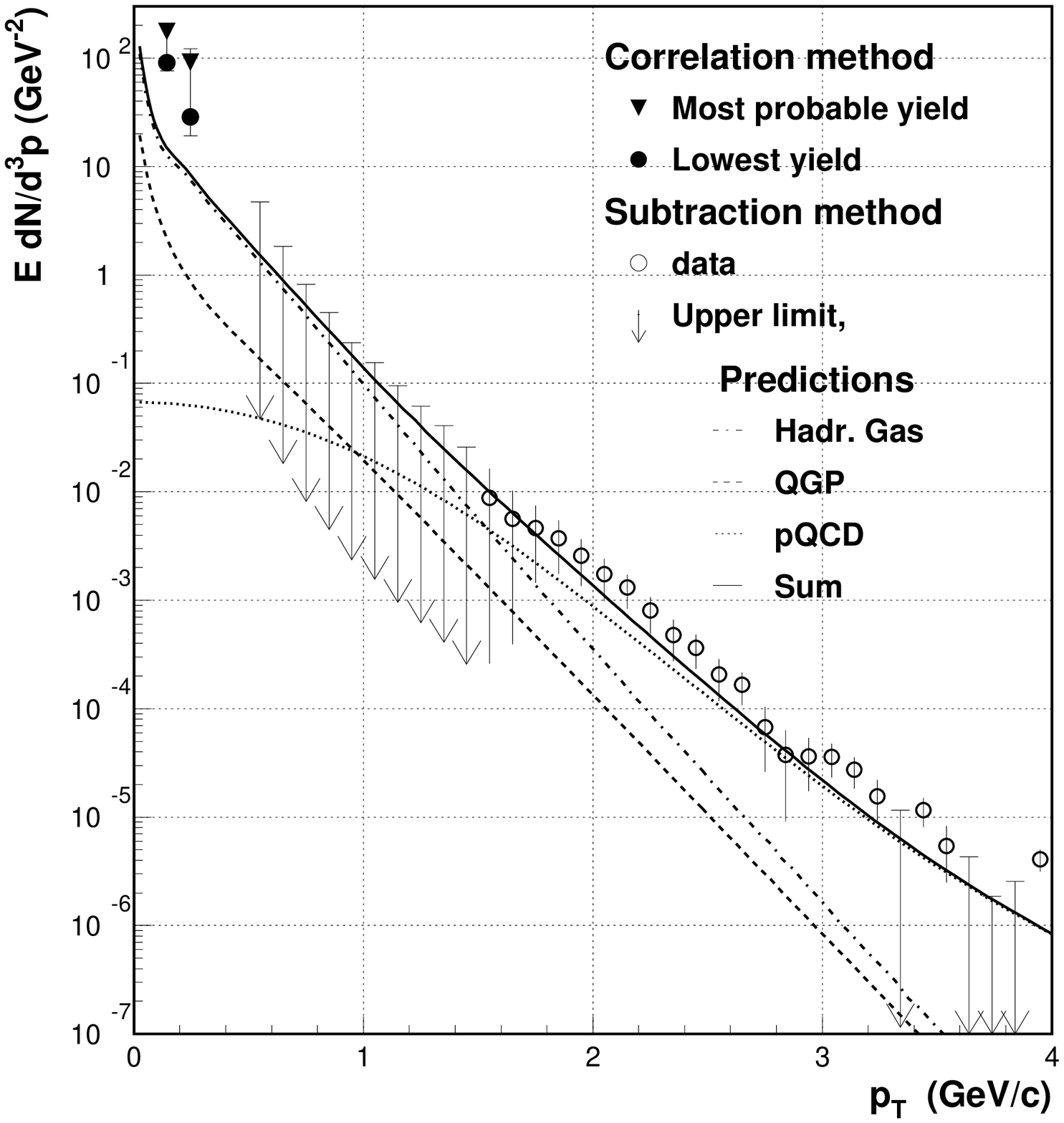}
  }
  \caption{Left plot: direct photon invariant correlation radius in
  Pb+Pb collisions at 158 AGeV (circles), compared to side (triangles) and long (boxes) charged
  pions correlation radii. Right plot: direct photon yield, extracted from
  correlation strength parameter, compared to theoretical predictions and
  yield, obtained withe statistical method \cite{wa98}.}
  \label{fig:wa98}
\end{figure}

The first measurement of the direct photon correlations in
ultrarelativistic heavy ion collisions was performed by WA98
collaboration \cite{wa98}. Its unique electromagnetic
calo\-ri\-meter, consisting of $4\cdot4\cdot40$~cm$^3$ lead glass
blocks, was situated at a distance of 21~m from the interaction
point. This provided an excellent opportunity to measure photon
pairs with very small relative momenta. The main difficulty in
this analysis was the separation between the apparatus effects
like cluster merging and splitting and the real physical
correlations, since the former strongly distort photon correlation
function at small $q$. This separation was done by introducing
cuts on a minimal dis\-ta\-nce between clusters and explo\-ring
the dependence of the final result on these cuts. How\-ever, at
$K_T \ge 0.5$ GeV photons with relative momenta $q<50$~MeV have so
small rela\-tive angle that it was not possible to perform such an
analysis any more and the in\-va\-riant correlation radius and the
strength parameter were measured only at small $K_T$, see the
Fig.~\ref{fig:wa98}. The photon invariant correlation radius was
close to the pion "side" and "long" radii and was considerably
above than the theoretical predictions \cite{Peressounko}. In
addition to the photon correlation radius, using the correlation
strength parameter, a proportion of the direct photons was
extracted and the direct photon yield was measured at a very small
$p_T$, where the other methods can not be applied. Since the
relation between the full three-dimensional and the
one-dimensional invariant correlation strength parameters
in\-vol\-ves $R_o$ -- radius which can not be estimated using
$R_{inv}$ -- a {\it lower} limit, cor\-re\-sponding to $R_o=0$ and
the most probable yield ($R_o=6$ fm) of direct photons was
extracted.

Presently there is a possibility to extract the direct photon
correlations in A+A collisions at RHIC energy with the ongoing
PHENIX and STAR experiments and at LHC with building ALICE
experiment. Experiment PHENIX has an electromagnetic calorimeter,
consisting of two parts, one of them is the same calorimeter used
in WA98 installed at a distance of 540~cm from the interaction
point while the rest of calorimeter has coarser granularity
$5.5\cdot5.5$~cm$^2$ and situated at a distance of 510 cm from the
interaction point. Although in PHENIX the calorimeter is 4 times
closer to the interaction point than one was in the WA98
experiment, this is compensated by smaller energies of photons
since PHENIX is a collider experiment so that it is able to access
photon pairs at small $q\sim 30$~MeV up to $K_T\sim 1$~GeV.
Experiment STAR is going to get advantage of their tracking
chamber and extract the direct photon correlations between
photons, one of which has converted into electron-positron pair on
a material of the detector and the second is detected with
calorimeter. The ALICE experiment at LHC will be even more
suitable for measuring of two-photon correlations. Its highly
granulated PHOS calorimeter made of $2\cdot2\cdot20$~cm$^3$
$PbW0_4$ crystals will be installed at a distance of 460~cm from
the interaction point and will have 4 times more channels per
solid angle and much better energy and position resolutions than
existing calorimeter in PHENIX. Si\-mu\-la\-tions show that PHOS
will be able to measure direct photon correlations up to $K_T\sim
2$ GeV.

To summarize, the direct photon correlations are very important
tool for exploring space-time evolution of the hot matter in
ultrarelativistic heavy ion collisions. Although there are plenty
of predictions, there is no agreement neither in absolute value of
the photon correlation radii nor even in the shape of their $K_T$
dependence. We considered several remarkable differences between
photon and hadron interferometry, related to the penetrating
nature and zero mass of a photon and the small yield of direct
photons. First results from the WA98 experiment demonstrated
possibility of measurement of the direct photon correlations in
ultrarelativistic heavy ion collisions. Ongoing experiments PHENIX
and STAR at RHIC as well as building experiment ALICE at LHC have
real possibility to measure the direct photon correlations so one
can expect more results soon.

\begin{theacknowledgments}
This work was supported by MPN of Russian Federation under the
grant NS-1885.2003.2 and by the INTAS grant No 04-83-4050.
\end{theacknowledgments}

\bibliographystyle{aipproc}   

\end{document}